\documentclass[aps,prb,twocolumn,superscriptaddress,floatfix,showpacs,citeautoscript,longbibliography,hyperlinks]{revtex4-2}
\usepackage{amsmath}
\usepackage{epstopdf}
\usepackage{amsmath}
\usepackage{amssymb}
\usepackage{amsfonts}
\usepackage[colorlinks=true,citecolor=blue]{hyperref}
\usepackage{graphicx,graphics}

\usepackage{float}
\usepackage[caption = false]{subfig}

\usepackage{xcolor}
\newcommand{\revision}[1]{{{#1}}}
\newcommand{\revisiontwo}[1]{{{#1}}}

\begin{document}
\title{Dynamical quantum phase transitions in strongly correlated\\ two-dimensional spin lattices following a quench}
\author{Fredrik Brange}
\affiliation{Department of Applied Physics, Aalto University, FI-00076 Aalto, Finland}
\author{Sebastiano Peotta}
\affiliation{Department of Applied Physics, Aalto University, FI-00076 Aalto, Finland}
\author{Christian Flindt}
\affiliation{Department of Applied Physics, Aalto University, FI-00076 Aalto, Finland}
\author{Teemu Ojanen} 
\affiliation{Computational Physics Laboratory, Physics Unit, Faculty of Engineering and
Natural Sciences, Tampere University, FI-33014 Tampere, Finland}
\affiliation{Helsinki Institute of Physics, FI-00014 University of Helsinki, Finland}

\begin{abstract}
Dynamical quantum phase transitions are at the forefront of current efforts to understand quantum matter out of equilibrium. Except for a few exactly solvable models, predictions of these critical phenomena typically rely on advanced numerical methods. However, those approaches are mostly restricted to one dimension, making investigations of two-dimensional systems highly challenging.
Here, we present evidence of dynamical quantum phase transitions in strongly correlated spin lattices in two dimensions. To this end, we apply our recently developed cumulant method [\href{https://journals.aps.org/prx/abstract/10.1103/PhysRevX.11.041018}{Phys.~Rev.~X {\bf 11}, 041018 (2021)}] to determine the zeros of the Loschmidt amplitude in the complex plane of time, and we predict the crossing points of the thermodynamic lines of zeros with the real-time axis, where dynamical quantum phase transitions occur. We find the critical times of a two-dimensional quantum Ising lattice and the \revision{XYZ} model with ferromagnetic or antiferromagnetic couplings. We also show how dynamical quantum phase transitions can be predicted by measuring the initial energy fluctuations, for example, in quantum simulators or other engineered quantum systems.
\end{abstract}

\maketitle

\section{Introduction}
Dynamical quantum phase transitions concern the critical behavior of many-body systems that are brought out of equilibrium by sudden parameter \revision{quenches~\cite{Heyl:2013,Zvyagin2016,Heyl:2018,Heyl:2019}}. Experimentally, isolated quantum many-body systems can now be prepared with specified initial conditions and made to evolve according to a Hamiltonian that can be designed with exquisite control~\cite{Bloch:2008,Zhang:2017,Jurcevic:2017,Flaeschner:2018,Guo:2019,Wang:2019}. Dynamical \revision{quantum} phase transitions may then occur at critical times, where the overlap -- the Loschmidt amplitude --  between the initial state and the time-evolved state vanishes, and the associated free-energy density becomes non-analytic in the thermodynamic limit~\cite{Karrasch2013,Andraschko2014,Vajna2014,Budich2016,Halimeh2017,Sadrzadeh:2021}. Dynamical quantum phase transitions extend the framework of equilibrium phase transitions~\cite{Chandler:1987,Goldenfeld:2018} to the time-evolution of quantum many-body systems \revision{with the aim to define phases of non-equilibrium quantum matter~\cite{Note1}.}

Theoretically, dynamical quantum phase transitions have been \revision{investigated} for exactly solvable problems~\cite{,Karrasch2013,Andraschko2014,Vajna2014,Budich2016,Halimeh2017,Sadrzadeh:2021}, starting with the Ising model in one dimension~\cite{Heyl:2013,Zvyagin2016,Heyl:2018}. Generally, however, the dynamics of interacting quantum many-body systems is rarely analytically tractable, and \revision{tensor-network methods have instead} been used to explore dynamical quantum phase transitions in one dimension~\cite{Pozsgay2013,Kriel2014,Sharma2015,ZaunerStauber2017,Homrighausen2017,Heyl:2018a,Kennes2018,Hagymasi:2019,Lacki2019,Huang2019,Halimeh2020}
The situation is even more challenging for systems in higher dimensions, such as two-dimensional quantum  \revision{lattices}~\cite{Heyl:2015,Schmitt:2015,Vajna2015,Bhattacharya:2017,Bhattacharya:2017b}. Here, progress has been hindered by the lack of efficient numerical approaches, and only a handful of interacting problems have been solved~\cite{Huang2019,Weidinger:2017,De_Nicola:2019,denicola:2021,hashizume:2020}. \revision{For this reason, the prediction of dynamical quantum phase transitions in two-dimensional systems with strong correlations has been identified as an important open problem, in particular for understanding the role of dimensionality for the critical dynamics~\cite{Heyl:2019}.}

\begin{figure}[b]
    \centering
    \includegraphics[width=0.94\columnwidth]{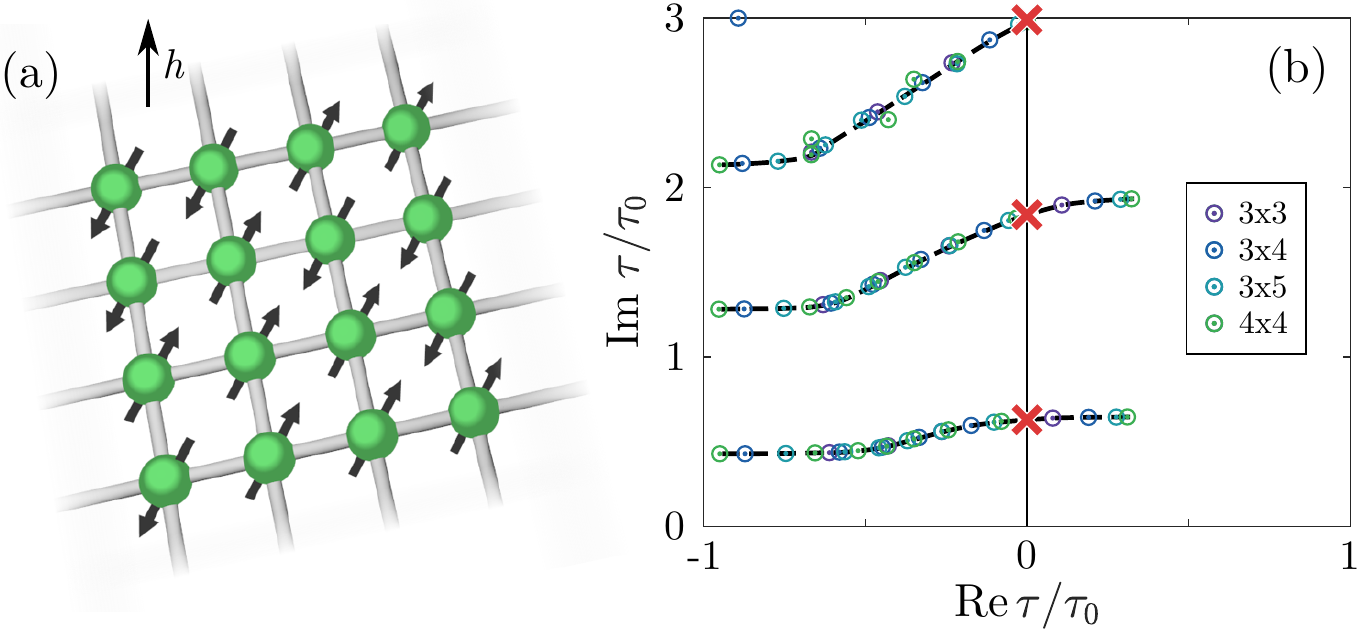}
    \captionsetup{justification=justified,singlelinecheck=false}
    \caption{Dynamical quantum phase transitions in a spin lattice. (a) Lattice of interacting spin-1/2 particles in a perpendicular magnetic field $h$. (b) Zeros of the Loschmidt amplitude in the complex plane of time for the quench $h_1=
    J\rightarrow h_2=5J$ with a ferromagnetic coupling, $J>0$, and $J_x = -1.5J$, $J_y = -0.5J$ and $J_z = -0.5J$ in the Hamiltonian~(\ref{eq:ham}) on different small lattices. The dashed lines are guides to the eye that help identify the crossings (red markers) with the imaginary axis corresponding to the critical times, where dynamical quantum phase transitions occur. The characteristic time scale $\tau_0$ is defined in Eq.~(\ref{eq:tau0}).}
    \label{Fig1}
\end{figure}

In this \revisiontwo{article}, we employ our recently developed  cumulant method  to predict dynamical quantum phase transitions in strongly-correlated two-dimensional spin lattices following a quench~\cite{Peotta:2021}. Our setup is illustrated in Fig.~\ref{Fig1}(a), showing a lattice of interacting spins in a magnetic field. We initialize the spins in the groundstate corresponding to one value of the field, and we investigate their time-evolution as we suddenly change the field strength. By promoting time to a complex variable in the spirit of Lee-Yang theory~\cite{Yang:1952,Lee:1952,Blythe:2003,Bena:2005,Deger:2018,Deger:2019,Deger:2020,Deger:2020b}, we use \revision{our cumulant method} to determine the zeros of the Loschmidt amplitude in Fig.~\ref{Fig1}(b), which allows us to construct the thermodynamic lines of zeros, whose crossing points with the imaginary axis \revision{(marked with red)} signal the critical times of the phase transitions. \revision{Importantly}, the dynamical quantum phase transitions can be predicted from the complex zeros obtained for small lattices instead of identifying cusps or other singularities in the rate function, which only become pronounced for large systems. \revision{Moreover, we can construct the full dynamical phase diagram, which consists of thermodynamic lines of zeros that separate different dynamical regions. In some cases, these lines separate an ordered phase from a disordered phase of an associated equilibrium system. Our work thereby establishes a connection between dynamical quantum phase transitions and related equilibrium phases of matter.}

\begin{figure*}[t]
	\centering
	\includegraphics[width=0.97\textwidth]{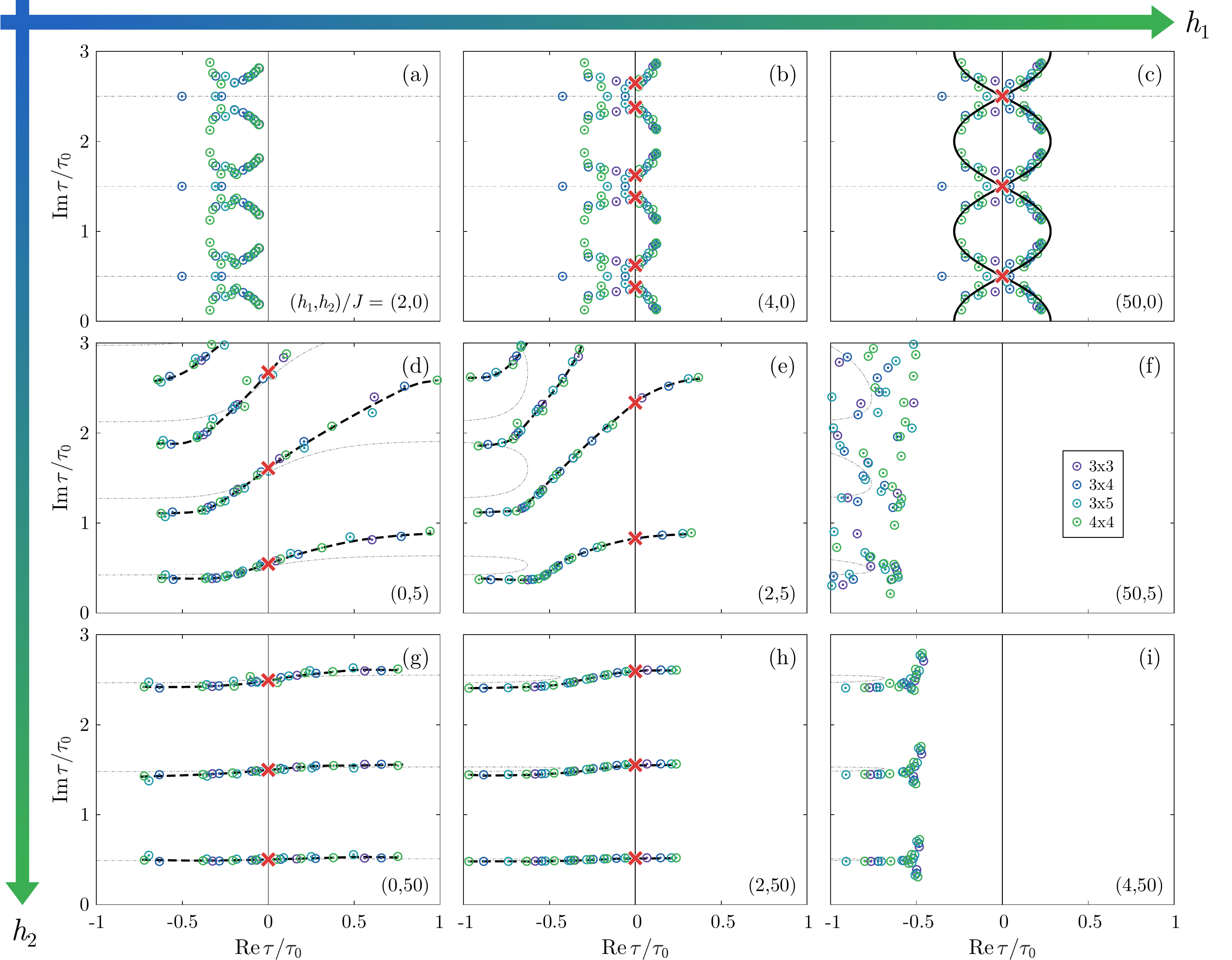}
	\captionsetup{justification=justified,singlelinecheck=false}
	\caption{Two-dimensional quantum Ising lattice. The zeros of the Loschmidt amplitude are shown for several quenches, $h_1\rightarrow h_2$, in quantum Ising lattices of different sizes. The dashed lines are guides to the eye that help identify the crossings with the imaginary axis, indicated by red markers, where dynamical quantum phase transitions occur in the thermodynamic limit. The solid line in panel (c) shows the thermodynamic lines of zeros for the classical Ising model found in Ref.~\cite{Saarloos:1984}. \revision{The gray dashed lines are the thermodynamic lines of zeros obtained in Ref.~\cite{Heyl:2013} for the quantum Ising model in one dimension.}}
	\label{Fig2}
\end{figure*}

\section{Spin lattice}
We consider a two-dimensional square lattice with interacting spin-1/2 particles in a perpendicular magnetic field $h$ as shown in Fig.~\ref{Fig1}(a). The system is described by the \revision{nearest-neighbor spin} Hamiltonian
\begin{equation}
\mathcal{\hat H}(h) = \sum_{\langle ij\rangle,\alpha} J_\alpha \hat S_i^\alpha \hat S_{j}^\alpha -h \sum_{i} \hat S_i^z,
\label{eq:ham}
\end{equation}
where $\hat S^\alpha_i$ is the spin-1/2 operator for each spin component, $\alpha = x,y,z$, on site $i$, and $J_\alpha$ denotes the exchange couplings. Here, the first sum runs over all nearest neighbor pairs $\langle ij\rangle$, and the second sum runs over all sites~$i$. To minimize edge effects, we impose periodic boundary conditions in both directions. We also define $J \equiv -(J_x+J_y)/2$, which is positive for ferromagnetic couplings and negative for antiferromagnetic couplings, and we note that the quantum Ising model is obtained by setting $J_y=J_z=0$.

\begin{figure*}[t]
    \centering
    \includegraphics[width=0.97\textwidth]{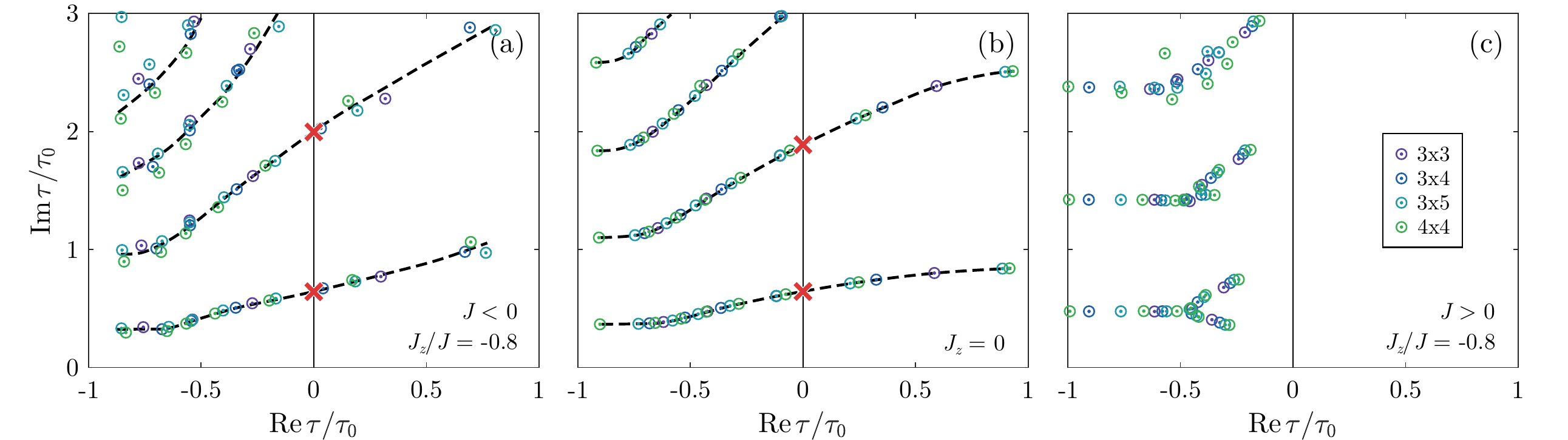}
    \captionsetup{justification=justified,singlelinecheck=false}
        \caption{\revision{XYZ} model. Zeros of the Loschmidt amplitude in the complex plane of time for the quench $h_1=|J|\rightarrow h_2=5|J|$ with (a) antiferromagnetic couplings $J_x=-1.5J$, $J_y=-0.5J$, $J_z=-0.8J$, $J<0$, (b) ferromagnetic or antiferromagnetic couplings $J_x=-1.5J$, $J_y=-0.5J$, $J_z=0$, $J\neq 0$, and (c) ferromagnetic couplings $J_x=-1.5J$, $J_y=-0.5J$, $J_z=-0.8J$, $J>0$. The dashed lines are guides to the eye, and the crossings with the imaginary axis are indicated by red markers.}
    \label{Fig3}
\end{figure*}

We now investigate the dynamical properties of the spins when prepared in the ground state, $|\Psi(t=0)\rangle=|\Psi_0\rangle$, corresponding to one value of the magnetic field, and then time-evolved in a different magnetic field for $t>0$. To this end, we consider the Loschmidt amplitude 
\begin{equation}
    \mathcal{Z}(it) = \langle \Psi_0 |e^{-it \mathcal{\hat H}(h_2)}|\Psi_0\rangle,
\end{equation}
where $h_1\rightarrow h_2$ is the quench of the magnetic field, and we have set $\hbar=1$. \revision{In the thermodynamic limit, $N \rightarrow \infty$, the rate function, $\lambda(t) = - \ln|\mathcal{Z}(it)|^2/N$, becomes non-analytic at the critical times, where the system exhibits a dynamical quantum phase transition. These phase transitions} can be understood as the zeros of $\mathcal{Z}(it)$ crossing the imaginary axis in the complex plane of $\tau=it$ for $N\rightarrow\infty$, where they form lines or areas. Thus, we aim to find the zeros of the Loschmidt amplitude for finite lattices, $\tau_k$, which can be factored as~\cite{Heyl:2013,Zvyagin2016,Heyl:2018,Yang:1952,Lee:1952,Blythe:2003,Bena:2005,Deger:2018,Deger:2019,Deger:2020,Deger:2020b}
\begin{equation}
\label{eq:factorized_Z}
 \mathcal{Z}(\tau=it) = e^{\alpha\tau}\prod_{k}\left(1-\tau/\tau_k\right)
\end{equation}
with $\alpha$ being a constant. We also define the Loschmidt cumulants, $\langle\!\langle \mathcal{\hat H}^n\rangle\!\rangle_\tau = (-1)^n \partial_\tau^n \ln \mathcal{Z}(\tau)$, which for $n>1$ can be expressed in terms of the zeros as 
\begin{equation}
    \langle\!\langle \mathcal{\hat H}^n\rangle\!\rangle_\tau = \sum_{k}\frac{(-1)^{n-1}(n-1)!}{(\tau_k-\tau)^n}.
\label{eq:loscumu}
\end{equation}
Generally, the Loschmidt cumulants are complex, however, at $\tau=0$ they reduce to the ordinary cumulants of the energy in the initial state with respect to the post-quench Hamiltonian, $\langle\!\langle \mathcal{\hat H}^n\rangle\!\rangle_0=\langle\!\langle E^n\rangle\!\rangle$. Moreover, by calculating several Loschmidt cumulants, we can invert Eq.~(\ref{eq:loscumu}) to find the zeros following Ref.~\cite{Peotta:2021}. \revisiontwo{(For calculating the Loschmidt cumulants, we use the Krylov subspace method described in Appendix~C of Ref.~\cite{Peotta:2021}.)} We then find the zeros in the vicinity of the chosen base point~$\tau$, and by using values of $\tau$ throughout the complex plane we can map out the zeros of the Loschmidt amplitude. 

\section{Quantum Ising lattice}
Figure~\ref{Fig2} shows zeros for a series of quenches in a quantum Ising lattice, which clearly exhibits dynamical quantum phase transitions at the critical times indicated by red markers. The zeros were obtained for small lattices, yet, we can identify the thermodynamic lines of zeros. The characteristic time scale
\begin{equation}
\tau_0 = \pi/\sqrt{J^2+h_2^2}
\label{eq:tau0}
\end{equation}
interpolates between $\tau_0 \simeq \pi/h_2$ for the field-dominated quenches in the bottom panels with $h_2\gg J$ and $\tau_0 \simeq \pi/J$ for the interaction-dominated quenches in the top row with $h_2\ll J$. Importantly, we observe dynamical quantum phase transitions \revision{only for quenches that cross the equilibrium critical point of about $h_c \simeq 3.04 J$~\cite{Blote:2002}. In panel~(g), the spins are confined to a plane, and they rotate in the strong magnetic field with crossings occurring at equidistant times spaced by~$\tau_0$. Interestingly, these results coincide with those of the one-dimensional quantum Ising model (indicated with gray dashed lines), showing that the dimensionaliy of the lattice plays only a little role for the field-dominated quenches. By contrast}, in panel~(c), all spins initially point in the $z$-direction, \mbox{$|\Psi_0\rangle = \bigotimes \limits_{i=1}^N|\!\uparrow^{(i)}_z\rangle=\bigotimes \limits_{i=1}^N\left(|\!\uparrow^{(i)}_x\rangle+|\!\downarrow^{(i)}_x\rangle\right)/\sqrt{2}$}, which is an equal superposition of all eigenstates of the postquench Hamiltonian with $h_2=0$. The Loschmidt amplitude is then given by the partition function of the classical Ising model, $\mathcal{Z}(it)=\mathrm{tr}\{e^{-it \mathcal{\hat H}(0)}\}/2^N$, with an imaginary inverse temperature, $\beta=it$ \cite{Heyl:2015}. In that case, the thermodynamic lines of zeros are known \cite{Saarloos:1984}, and they are shown with solid lines in panel~(c) as a check of our results. \revision{For the two-dimensional lattice, the thermodynamic lines of zero reach the real axis, corresponding to the thermal phase transition of the classical Ising model in two dimensions. This is different from the one-dimensional model, where the zeros stay off from the real axis due to the absence of spontaneous symmetry-breaking. Interestingly, panel~(c) illustrates how a dynamical quantum phase transition occurs as we cross the same thermodynamic line of zeros that separates the ordered and the disordered phases of the classical Ising model in two dimensions. For quenches with all energy scales, $h_1, h_2, J$, being of the same order, as in panel~(e), the spins are strongly correlated, and our methodology is needed to predict the critical times. Altogether, Fig.~\ref{Fig2} provides a detailed dynamical phase diagram of the two-dimensional quantum Ising model following a field quench.}

\section{\revision{XYZ model}}
The quantum Ising lattice is a prime example of a two-dimensional quantum many-body system that exhibits dynamical quantum phase transitions. However, our method has a much broader scope, and we now go to the paradigmatic \revision{XYZ} model, which, in contrast to the quantum Ising model, exhibits different physics for ferromagnetic and antiferromagnetic couplings. To avoid artificial lattice frustration in the latter case, we use anti-periodic boundary conditions along directions with an odd number of sites for antiferromagnetic couplings. Figure~\ref{Fig3} shows zeros of the Loschmidt amplitude for quenches in small lattices with antiferromagnetic or ferromagnetic couplings. Dynamical phase transitions clearly occur with antiferromagnetic couplings in panel (a). By contrast, as $J_z$ evolves from being positive in Fig.~\ref{Fig1}(b) to being zero and negative in panels (b) and (c) of Fig.~\ref{Fig3}, the dynamical quantum phase transitions eventually go away, and the zeros stay off from the imaginary axis in panel~(c).

\section{Experimental signatures}
\begin{figure*}
	\centering
	\includegraphics[width=0.97\textwidth]{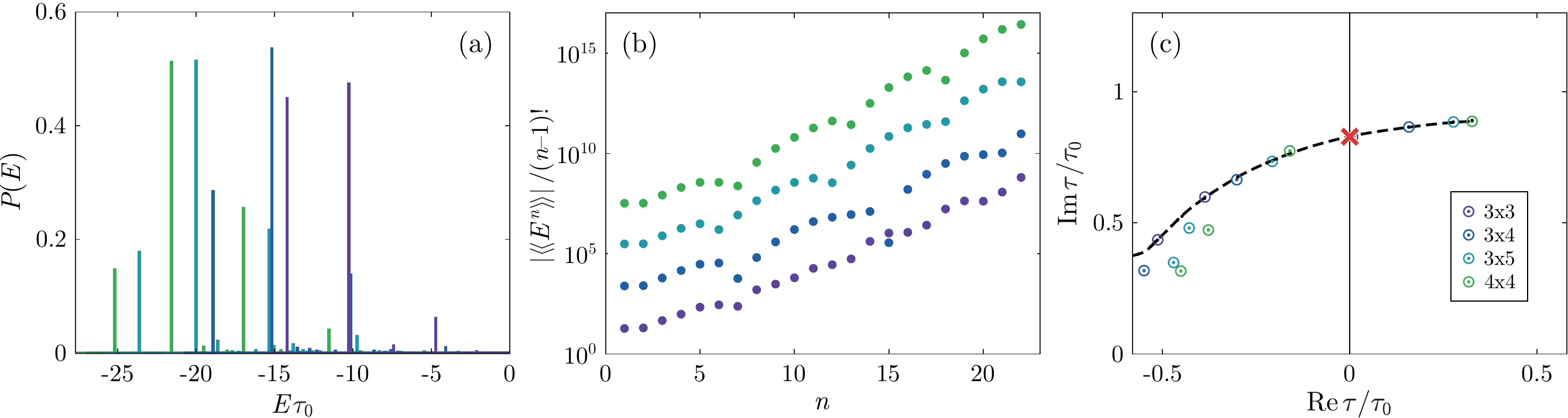}
	\captionsetup{justification=justified,singlelinecheck=false}
	\caption{Experimental signatures. (a) The distribution of the energy, given by the post-quench Hamiltonian, in the initial state, obtained from $10^6$ Monte Carlo simulations for the quench $h_1=2J \rightarrow h_2 =5J$ and $J_x=-1.5J$, $J_y=-0.5J$ with $J>0$ and $J_z=0$ in Eq.~(\ref{eq:ham}) on different lattice sizes. (b) High cumulants of the energy distributions, displaced vertically for the sake of clarity. (c)~Zeros of the Loschmidt amplitude obtained from the cumulants in panel (b) with the dashed line being a guide to the eye that helps identify the crossing (red marker) with the imaginary axis, where a dynamical phase transition occurs.}
	\label{Fig4}
\end{figure*}

Finally, we show how dynamical quantum phase transitions can be predicted from measurements of the energy fluctuations in the initial state only. At $\tau=0$, the Loschmidt cumulants reduce to the ordinary cumulants of the postquench Hamiltonian with respect to the initial state. Thus, by repeatedly preparing the system in the state $|\Psi_0\rangle$ and measuring the energy given by the postquench Hamiltonian, one can construct the energy distribution and extract the corresponding cumulants for $\tau=0$. (By contrast, the results presented in Figs.~\ref{Fig2} and~\ref{Fig3} were obtained by calculating Loschmidt cumulants for many values of $\tau$ throughout the complex plane.)  This approach is illustrated in Fig.~\ref{Fig4}, where we mimic experimental data by simulating $10^6$ measurements of the initial energy and construct the distributions in panel~(a) for a few different lattice sizes. In panel~(b), we show the cumulants of the distributions and in panel~(c) we show the zeros of the Loschmidt amplitude that we extract from the energy cumulants by inverting Eq.~(\ref{eq:loscumu}) for the zeros with $\tau=0$. These results demonstrate how the first dynamical phase transition can be predicted from the initial energy fluctuations, which one could hope to measure, for example, on near-term quantum simulators~\cite{Lin:2021,Sun:2021} or in other engineered quantum systems such as small atomic structures on surfaces~\cite{Choi:2019} or ultracold atoms in optical lattices~\cite{Eckardt:2017}. \revision{Cumulants of order 15 and higher have been measured for electron transport through quantum dots~\cite{Flindt:2009} and metallic islands~\cite{Maisi:2014}, and the corresponding  Lee-Yang zeros were extracted in Ref.~\cite{Brandner_2017}, showing that this approach to dynamical phase transistions is experimentally feasible.} 

\section{Conclusions and outlook}
We have presented evidence of dynamical quantum phase transitions in two-dimensional quantum Ising lattices and in the \revision{XYZ} model with ferromagnetic or antiferromagnetic couplings. To this end, we have employed a systematic strategy, whereby we determine the zeros of the Loschmidt amplitude in the complex plane of time, allowing us to identify the thermodynamic lines of zeros and their crossings with the real-time axis, where dynamical phase transitions occur. This approach is also of experimental relevance, as it makes it possible to predict the first critical time of a dynamical quantum phase transition by measuring the energy fluctuations in the initial state. Our work opens avenues for a wide range of applications, and we conclude with an outlook on possible directions for future developments: an immediate and interesting extension of the present work would be to consider frustrated spin models. So far, we have only considered spin lattices, however, quantum many-body systems involving interacting fermions or bosons can also be treated. \revision{Furthermore, it would be interesting to investigate the two-dimensional quantum Ising model with long-range interactions to see if connections to other notions of dynamical quantum phase transitions can be established as in Ref.~\cite{Zunkovic:2018}.}

\section{Acknowledgements} 
We thank A.~Deger for useful discussions. The work was supported by the Academy of Finland through the Finnish Centre of Excellence in Quantum Technology (Projects No. 312057 and 312299) and Projects No.~336369, 330384, 331094, and 331737. T.~O.~acknowledges support from Helsinki Institute of Physics.


%

\end{document}